\documentclass{svjour2}                    
\smartqed  
\usepackage{graphicx}
\begin{document}

\title{Electromagnetic waves from neutron stars and black holes driven by polar gravitational perturbations}


\titlerunning{EM waves from NSs and BHs}        

\author{Hajime Sotani, Kostas D. Kokkotas,  Pablo Laguna, Carlos F. Sopuerta}


\institute{H. Sotani \at
              Yukawa Institute for Theoretical Physics, Kyoto University, Kyoto 606-8502, Japan \\
              \email{sotani@yukawa.kyoto-u.ac.jp}             
           \and
           K. D. Kokkotas \at
              Theoretical Astrophysics, IAAT, Eberhard Karls University of T\"ubingen,  72076, T\"ubingen, Germany
           \and
           P. Laguna \at
              Center for Relativistic Astrophysics \& School of Physics, Georgia Institute of Technology, Atlanta, GA 30332, USA
           \and
           C. F. Sopuerta \at
              Institut de Ci\`encies de l'Espai (CSIC-IEEC), Campus UAB, Facultat de Ci\`encies, Torre C5 parells, Bellaterra, 08193 Barcelona, Spain
}

\date{Received: date / Accepted: date}

\maketitle

\begin{abstract}
Neutron stars and black holes are the most compact astrophysical objects we can think of and as a consequence they
are the main sources of gravitational waves.  There are many astrophysically relevant scenarios in which these 
objects are immersed in or endowed with strong magnetic fields, in such a way that gravitational perturbations can 
couple to electromagnetic ones and can potentially trigger synergistic electromagnetic signatures.  
In a recent paper we derived the main equations for gravito-electromagnetic perturbations and studied in detail the
case of polar electromagnetic perturbations driven by axial gravitational perturbations.  In this paper we deal with
the case of axial electromagnetic perturbations driven by polar black-hole or neutron stars oscillations, in which the energy 
emitted in case is considerably larger than in the previous case. In the case of neutron stars the phenomenon lasts 
considerably longer since the fluid acts as an energy reservoir that shakes the magnetic field for a timescale of 
the order of secs.
\end{abstract}

\section{Introduction}
\label{sec:Intro}
Gravitational wave observations will enable an unprecedented view of the Cosmos thanks to interferometric detectors such as the forthcoming second-generation ground-based detectors LIGO, Virgo, and KAGRA~\cite{LIGO,VIRGO,KAGRA}, third-generation future ones like the Einstein Telescope~\cite{ET}, and space-based ones like eLISA and DECIGO~\cite{eLISA,DECIGO}.  In addition, there are also good prospect of detection of gravitational waves using several Pulsar Timing Arrays that are organized in the International Pulsar Timing Array collaboration~\cite{IPTA}.  When the second generation of ground-based detectors becomes fully operational and reaches the designed sensitivity, these instruments will be truly tools of astronomical discovery. For compact objects (neutron stars and black holes), we will be able to determine from gravitational-wave observations properties such as their masses, radius, rotation rates, and, for the case of neutron stars, the structure of matter itself (i.e. equation of state)~\cite{AK1998,KAA2001,Sotani2004,Erich2011,Sotani2011a}. Another remarkable capability of gravitational-wave observations is their potential to validate whether General Relativity is indeed the correct classical theory of gravity and to put constraints to alternative theories~\cite{SK2004,S2009,YYT2012}. In summary, the detection and characterization of gravitational waves will provide insights on a plethora of physical phenomena, insights that will complement our knowledge obtained from particle accelerators and electromagnetic observations. 

Processes in compact objects associated with strong magnetic fields are an example of physical phenomena for which only multi-messenger observations provide a complete picture. For instance, in the case of the giant flares in soft gamma-ray repeaters, evidence points to magnetars (strongly magnetized neutron stars) as the objects where these events takes place. 
Moreover, in the three giant flares detected so far, theoretical studies~\cite{L2007,SKS2007,CBK2009,CSF2009,AGS2009,S2011,HoLevin11,GCFES2012,CK2012,SNIO2012a,SNIO2012b,HoLevin12} suggest that the quasi-periodic oscillations frequencies, measured during the afterglow of the flare activity~\cite{WS2006}, are associated with crustal oscillations in the neutron star and/or oscillations in the magnetic field of the neutron star. Furthermore, recent  studies~\cite{A2011} have speculated that the flare activity could couple with the dynamical gravitational fields of the star to trigger the production of gravitational waves.
In addition, the importance of electromagnetic counterparts to gravitational-wave signals has been also discussed in the context of recent numerical simulations of binary black hole systems~\cite{PL2010,PT2010}, of collapsing hypermassive neutron stars~\cite{LP2012}, and of binary neutron star systems~\cite{PL2013a,PL2013b}.

Further studies are needed  to investigate the conditions under which electromagnetic and gravitational perturbations couple to generate emission of radiation through both channels. We have taken a first step in recent work~\cite{SKLS2013} (Paper I). We studied the coupling of axial gravitational and electromagnetic perturbations for both, black holes and neutron stars, neglecting the back reaction of the electromagnetic waves since the energy budget in the system is dominated by gravity. Our study found that the emitted energy in electromagnetic 
waves driven by the gravitational waves is proportional to not only the emitted energy in gravitational 
waves but also to the square of the strength of the magnetic field of the central object. That is, $E_{\rm EM}/E_{\rm GW} = \alpha (B/10^{15} {\rm G})^2$, where $B$ is the magnetic field 
strength. For the case of a black hole background it was found that $\alpha \approx 8\times 10^{-6}$, while for neutron stars this factor depends on the compactness of the neutron star.  For a compactness of $M/R = 0.162$ we obtained $\alpha \approx 1.61\times 10^{-5}$ while for a compactness of $M/R = 0.237$ the result was $\alpha \approx 4.37\times 10^{-6}$. 
Additionally, it is suggested that there can be resonances between the two types of waves under certain conditions \cite{KT2013}.

In this paper, we investigate electromagnetic emission driven by polar gravitational perturbations. Unlike axial gravitational perturbations, the expectation is that polar perturbations are able to drive longer emission in both gravitational and electromagnetic waves. Here again, we ignore back-reaction effects from the magnetic field on the spacetime geometry.  Following the previous study in Paper I, we consider both the cases of black holes and neutron stars.  We examine in particular the dependence of the gravitational-electromagnetic coupling for neutron stars on the stellar mass and the equation of state. 
Then, we find that $\alpha=1.78\times 10^{-5}$ for a black hole background, a value roughly twice the one found in Paper I for electromagnetic waves driven by the axial gravitational waves. On the other hand, we also find that $\alpha$, for a neutron star background, can be written as a function of ${\cal \chi} \equiv (1-2M/R)^{-1}(M/R^3)^{-1}$ as $\alpha \times 10^7 = 0.922 - 0.903({\cal \chi}/{500}) + 0.552 ({\cal \chi}/{500})^2$, where $M$ an $R$ denote the stellar mass and radius.

This paper is organized as follows. In the next section (Section~\ref{sec:Background}), we briefly describe the equilibrium configurations that provide the gravitational background (spacetime background geometry) as well as the the perturbative equations. In Section~\ref{sec:Results}, we show and discuss the numerical results for both the black hole and neutron star cases. We make conclusions in Section~\ref{sec:conclusion}. Throughout this paper we adopt geometric units, $c=G=1$, where $c$ and $G$ denote the speed of light and the gravitational constant, respectively, and use the metric signature $(-,+,+,+)$.

\section{Spacetime Background and Perturbation Equations}
\label{sec:Background}

Since the systems we are considering involve dynamical gravitational and electromagnetic fields and, for the case of neutron stars, fluids, the equations to be solved are the Einstein-Maxwell equations, coupled to the relativistic hydrodynamic equations for the neutron star case. In general, the presence of magnetic fields causes deformations in compact objects due mostly to non-radial magnetic pressure. However, for astrophysically relevant situations, the magnetic energy is much smaller than the gravitational binding energy of the compact object. For instance, in magnetars, with expected magnetic field strengths of the order of $\sim 10^{16}$G, the ratio of the magnetic energy to the gravitational binding energy  is ${\cal E}_{\rm B}/{\cal E}_{\rm G} =  10^{-4}\,(B/10^{16}\,{\rm G})^2$. Therefore, it is then safe to neglect the back-reaction of the magnetic fields on the background spacetime metric geometry. That is, the coupling between gravitational and electromagnetic perturbations is such that gravitational perturbations influence the dynamics of electromagnetic perturbations but not vice versa. For simplicity, we assume non-rotating black holes and neutron stars; thus, we use the Schwarzschild metric for the exterior of the black hole and the neutron star,  and the Tolman-Oppenheimer-Volkoff (TOV) solution in the interior of neutron stars (see, e.g.~\cite{BFS85}).  This solution can be written as 
\begin{equation}
  ds^2=-e^{\nu}dt^2+e^{\lambda}dr^2 +r^2(d\theta^2+\sin^2\theta d\phi^2)\,, 
  \label{eq:Schw}
\end{equation}
where the metric functions $\nu(r)$ and $\lambda(r)$, in the interior of a neutron star, are determined by the TOV equations. In the exterior, 
as in the case of a black hole, they are determined by the standard Schwarzschild solution: $e^{-\lambda}=e^{\nu}=1-2M/r$.

Regarding the magnetic field, we consider only dipole configurations.  Detailed expressions of the dipole magnetic fields in the interior of the neutron star and the exterior of the black hole and neutron star can be found in Paper I.  In the same way, in this paper we continue using the ideal MagnetoHydroDynamic (MHD) approximation to model the neutron star.


Given the background equilibrium configurations described above, we add small perturbations to both the gravitational and electromagnetic fields. The equations governing these small perturbations are obtained by linearizing the Einstein-Maxwell equations (coupled to the fluid stress-energy tensor variables in the case of neutron stars). 
The energy stored in the gravitational perturbations can be significantly larger than that in the electromagnetic perturbations, since the gravitational binding energy of the equilibrium configurations is typically a few orders of magnitude larger than the energy of the magnetic field.  
Following this argument we then neglect the back-reaction of the electromagnetic perturbations on the gravitational ones in this paper, in the same way as in Paper I. 


The detailed form of the coupled system of perturbation equations for both the electromagnetic and gravitational fields are described and presented in detail in Paper I. There we have shown that the dipole ``electric (polar) type" electromagnetic perturbations are driven by axial quadrupole gravitational perturbations, while dipole ``magnetic (axial) type" electromagnetic perturbations are driven by polar quadrupole gravitational perturbations. The first case has already been studied in Paper I while the details of the second case are presented in this paper.  This second case is considerably more involved than the first one, especially for the case of neutron stars. Moreover, the interaction between the two fields lasts longer due to the presence of the fluid which acts as an energy reservoir. As a result, the energy emitted in the form of electromagnetic waves is considerably larger than in the case studied in Paper I. 
The perturbation equations for the second case, i.e. for dipole magnetic-type (axial) electromagnetic perturbations driven by polar quadrupole gravitational perturbations, are summarized in Appendix.

\section{Numerical Results}
\label{sec:Results}

In this section we present results of numerical computations of the generation of dipole ``magnetic'' (axial) type electromagnetic perturbations 
driven by ``electric'' (polar) quadrupole gravitational perturbations both for black holes and neutron stars.

\subsection{Black Hole Background}
\label{sec:result1}

First, we consider the case of a black hole immersed in a dipole magnetic field. This case is simpler than the neutron star case because it involves finding solutions to only the gravitational and electromagnetic perturbation equations. As it was pointed out in Paper I, there are issues with the background dipole magnetic field near the event horizon. The magnetic field diverges logarithmically as one approaches the event horizon, more specifically $B \propto\mu_d\, \ln(1-2\,M/r)$, with $\mu_d$ being the magnetic dipole moment that an observer at infinity sees~\cite{Wasserman1983}. It is important to remark that the presence of a magnetic field around the black hole is not incompatible with the fact that an isolated black hole can not sustain a magnetic field due to the no-hair theorem of General Relativity.  What we are attempting to model here is a black hole in an astrophysical environment, such as the case of a black hole with an accretion disk.  In this sense, we envision the dipole magnetic field in the vicinity of the black hole as being anchored to the accretion disk~\cite{TNM2011,MTB2012}.  In practice, as we did in Paper I, 
we adopt the condition that $B(r)=B(6M)$ for $r\le 6M$, because $B=$ const is the valid solution near the horizon, where we have only freedom in deciding where the transition between the internal constant and external dipole solutions occurs. We especially decide that $r=6M$ as the transition position, which is motivated by the location, in the case of non-rotating black holes, of the innermost stable circular orbit for test massive particles at $r=6M$.
We identify the magnetic dipole moment $\mu_d$ with the normalized magnetic field strength $B_{15}\equiv B_p/(10^{15}\, {\rm G})$, where $B_p$ is the field strength determined at $r=6M$ and $\theta=0$.
We remark that although we have adopted $r=6M$ for the transition position, this selection does not affect significantly the results in this paper provided the transition position is larger than $r=3M$.

In order to solve numerically the perturbative equations we use the same techniques as in Paper I (see~\cite{Sotani2006} for more details), namely a Crank-Nicholson method~\cite{Teukolsky2000} on a mesh with uniform grid-spacing in the radial tortoise coordinate (defined as $r_* = r+2M\ln(r/2M-1)$) $\Delta r_* = 0.1M$ (see~\cite{Sotani2006} for the convergence tests of the results with respect to the grid spacing).  The time step is chosen such that $\Delta t = \Delta r_*/2$.  We choose vanishing initial electromagnetic perturbations, i.e., $\Phi_{10}=\partial_t\Phi_{10}=0$, but an ingoing Gaussian wave packet for the gravitational perturbation $Z_{20}$. That is, $Z_{20}(r_*)\propto \exp[-(r_*-r_0)^2/\sigma^2]$ and $\partial_t Z_{20}=\partial_{r_*}Z_{20}$, where $r_0\simeq 2000M$ and $\sigma=0.5M\,$.

In the course of the evolutions, we compute the luminosity emitted in gravitational waves, $L_{\rm GW}$, and in electromagnetic waves, $L_{\rm EM}$, using the well-known expressions~\cite{CPM-I}
\begin{equation}
  L_{\rm GW} = \frac{\left|\partial_t Z_{20}\right|^2}{384\pi} \ \ {\rm and}\ \ 
  L_{\rm EM} = \frac{\left|\partial_t \Phi_{10}\right|^2}{2\pi} \label{eq:LEM} \, , 
\end{equation}
where it is understood that we consider only the dominant modes, $l=1$ for the electromagnetic perturbations and $l=2$ for the gravitational perturbations.  The total energy emitted in electromagnetic and gravitational waves is obtained by integrating in time these luminosities.

As it happened in the case of axial gravitational perturbations driving ``electric-type" electromagnetic waves, considered in Paper I, we find that the electromagnetic-wave and gravitational-wave energies emitted, $E_{\rm EM}$ and $E_{\rm GW}$ respectively, satisfy the following relationship
\begin{equation}
  E_{\rm EM} = \alpha B_{15}^2E_{\rm GW}, \label{eq:E}
\end{equation}
with $\alpha$ a proportionality constant depending on the magnetic field distribution. 

\begin{figure*}[htbp]
\begin{center}
\begin{tabular}{cc}
\includegraphics[width=0.45\textwidth]{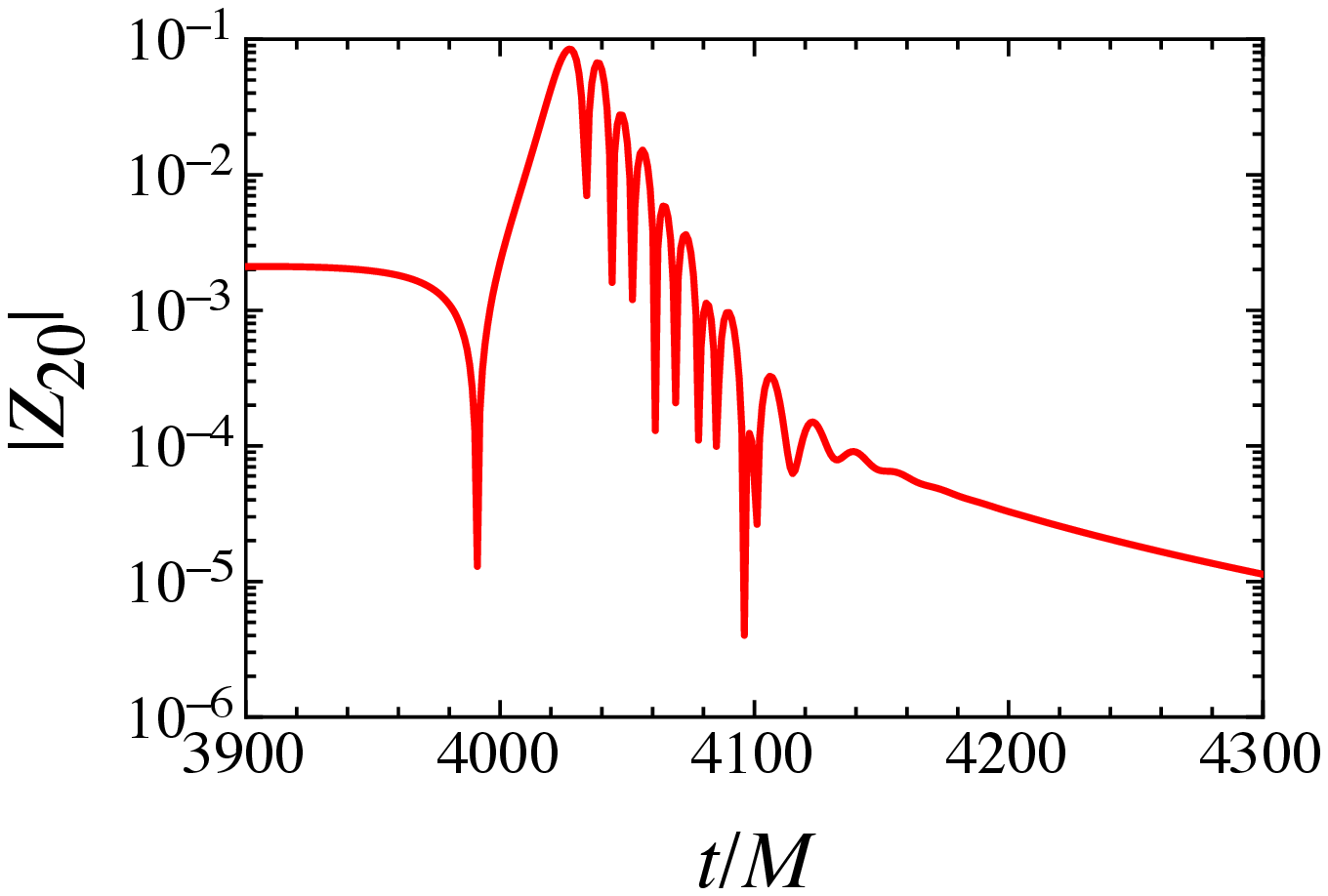} 
\includegraphics[width=0.45\textwidth]{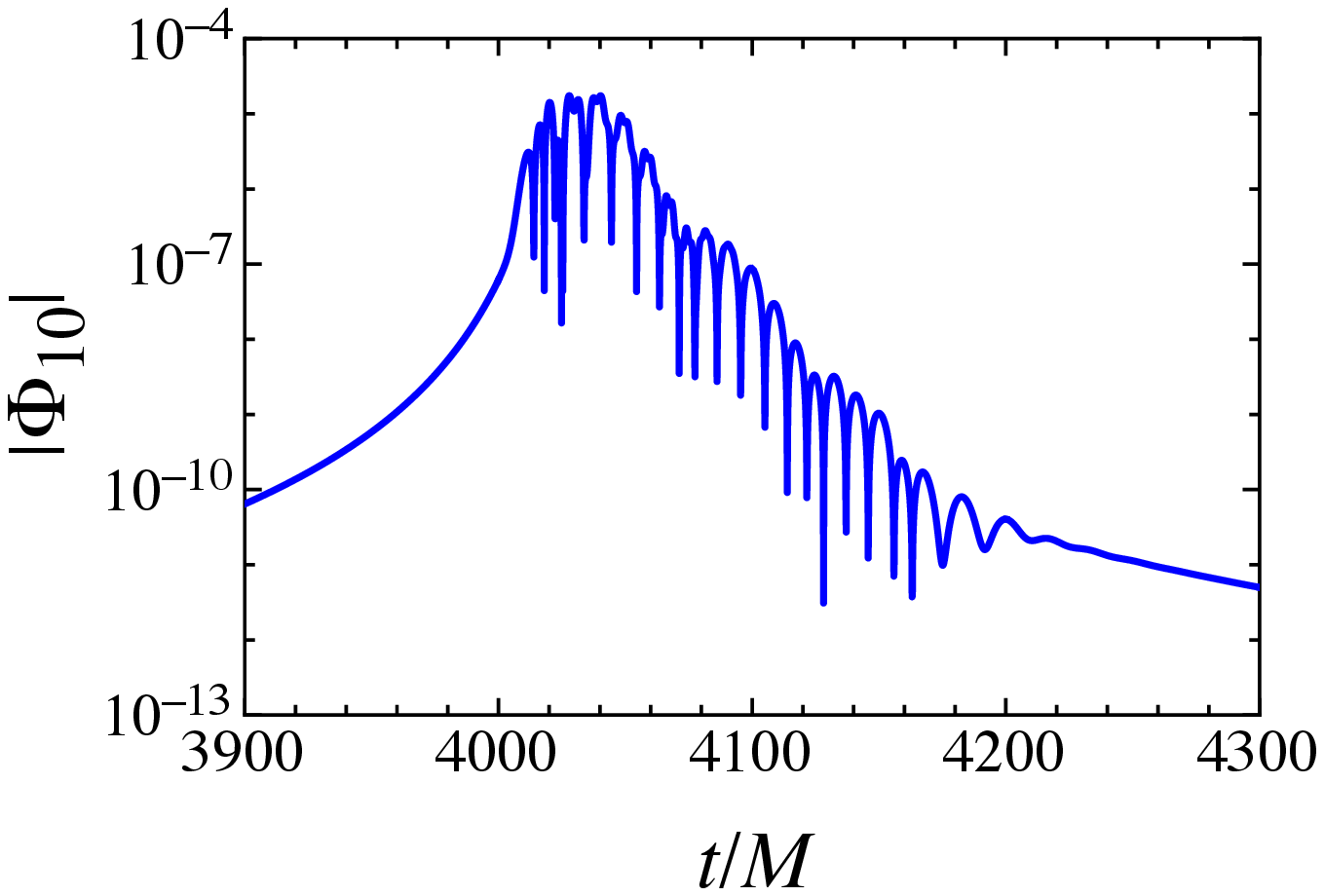}
\end{tabular}
\end{center}
\caption{
Absolute value of the polar gravitational wave $Z_{20}$ (left panel) and axial electromagnetic wave $\Psi_{10}$ (right panel) as observed at $r=2000\,M$.
}
\label{fig:GW}
\end{figure*}

In what follows, we present results of numerical simulations where the strength of magnetic field has been fixed to be $B_p=10^{15}$G, i.e. $B_{15} = 1$. In Figure~\ref{fig:GW}, we show the amplitude of the polar gravitational wave $Z_{20}$ (left panel) and of the axial electromagnetic wave $\Phi_{10}$ (right panel) as observed at $r=2000\,M$, where the amplitude of the waves has been normalized so that the emitted energy in gravitational waves satisfies $E_{\rm GW}\approx 2.2\times 10^{49}(M/50M_\odot)$ ergs, which is of the order of the typical energy radiated from physical phenomena associated with the compact objects. With this normalization, the resulting energy emitted in electromagnetic waves is $E_{\rm EM}\approx 3.9\times 10^{44}(M/50M_\odot)$ ergs. From these energies emitted in gravitational and electromagnetic waves, and using Eq.~(\ref{eq:E}), we find that
$\alpha=1.78\times 10^{-5}$, a value roughly twice the one found in Paper I for electromagnetic waves driven by the axial gravitational waves.  This is a consequence of the different coupling with the background magnetic fields.

\begin{figure}[htbp]
\begin{center}
\includegraphics[width=0.45\textwidth]{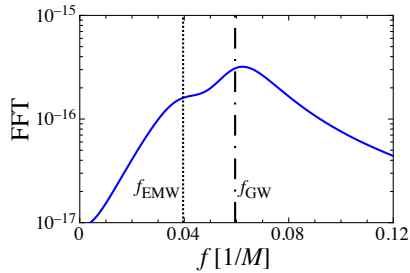}
\caption{
FFT of the axial electromagnetic perturbation $\Phi_{10}$ corresponding to the waveform in the right panel of Fig.~\ref{fig:GW}. The two vertical lines show the frequencies of the quasi-normal modes for the ``magnetic'' (axial) $l=1$ electromagnetic waves (dashed line) and for the ``electric'' (polar) $l=2$ gravitational waves (dot-dash line).
}
\label{fig:FFT-BH}
\end{center}
\end{figure}

Another remarkable fact that can be seen in the right panel of Figure~\ref{fig:GW} is that the electromagnetic waveform exhibits two different modes of oscillation. One of them, with shorter wavelength, seems to mimic the oscillations of the gravitational waves.  The other mode of oscillations, with longer wavelength, is only present in the electromagnetic waves. The presence of these two modes is more evident in Fig.~\ref{fig:FFT-BH}, where we show the fast Fourier transform (FFT) of 
$\Phi_{10}$.  The two bumps signalled by the vertical lines are the frequencies $f_{\rm EMW}=0.0396/M$ and $f_{\rm GW}=0.0595/M$  of the Schwarzschild  black hole quasi normal modes for $l=1$ electromagnetic and $l=2$ gravitational waves~\cite{CPM-I}. Consistent with our observation in terms of wavelengths, the electromagnetic mode has a smaller frequency than the gravitational one.

\begin{figure}[htbp]
\begin{center}
\includegraphics[width=0.45\textwidth]{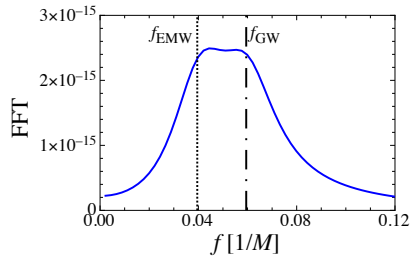}
\caption{
FFT of the electromagnetic perturbation $\Phi_{10}$ from the results of Paper I of the study of electromagnetic waves driven by axial gravitational perturbations. Two vertical lines denote the frequencies of the quasi-normal modes for the ``electric'' (polar) $l=1$ electromagnetic waves (dashed line) and for the ``magnetic'' (axial) $l=2$ gravitational waves (dot-dash line).
}
\label{fig:FFT-BHaxial}
\end{center}
\end{figure}

In paper I, we also found two quasi-normal frequencies in the electromagnetic waves.  The results found for electromagnetic waves driven by axial gravitational perturbations are reproduced in Figure~\ref{fig:FFT-BHaxial}. Notice, however, that the shape of the spectra is different. The reason for this is the specific nature of the gravito-electromagnetic coupling in each case. This imprint of the quasi-normal modes of gravitational waves in the spectra can provide clues about the gravitational-wave emission mechanisms from observations of the electromagnetic waves. However, as it was pointed out in Paper I, it might be quite difficult to detect such electromagnetic waves because at these low frequencies the electromagnetic waves could be easily coupled and/or absorbed by the interstellar medium as well as by the coupling with a putative accretion disc around the black hole.
On the other hand, there is also a proposal to build a lunar-based very low frequency radio array, which could provide a way of observing such low-frequency electromagnetic waves with frequencies of the order of $\sim 50$kHz if they approach the Earth \cite{luna01,luna02}.

%
%

\subsection{Neutron Star Background}
\label{sec:result2}

In the case of neutron stars, the prescription of initial conditions involves specifying data for the metric perturbations $F_{20}(r)$ and $S_{20}(r)$ (see Paper I), as well as for the fluid perturbation $H_{20}(r)$. These functions cannot be chosen freely but they must satisfy the Hamiltonian constraint.  The procedure that we apply to solve the equations follows the approach adopted by Allen et al~\cite{Allen1998}, as described in Paper I:   We specify first the variables $H_{20}$ and $S_{20}$ by setting $S_{20}(r)=0$ and $H_{20}(r) = H_a\,C_s^2(r/R)^2\cos(\pi r/2R)$, where $H_a$ is an arbitrary amplitude, $R$ is the stellar radius, and $C_s$ is the fluid sound speed.  And second, we solve the Hamiltonian constraint for the metric perturbation $F_{20}$, which completes the initial data construction. These initial conditions are inspired by the weak-field limit, where $S_{lm} \rightarrow 0$ as the star becomes Newtonian~\cite{T1969}.  In the approach of Allen et al~\cite{Allen1998} the evolution of our variables is by three coupled wave-type equations, two for the metric perturbations and one for the fluid perturbations.

We model the neutron star as a polytrope with equation of state $p = K\rho^\Gamma$.  In this paper we use the same parameters $(K,\Gamma)$ that we reported in Table I of Paper I. In order to establish comparisons among different stellar models, we fix the amplitude $H_a$ so that the initial energy produced by the density perturbations is approximately $0.01\,M_\odot$.  Following our notation, the initial energy is calculated via 
\begin{equation}
 E_0 = \frac{1}{2}\int \left(\frac{\delta p_{lm}}{\rho} - re^{-\nu}S_{lm} - \frac{F_{lm}}{r}\right)\delta\rho_{lm}^*d^3x,
\end{equation}
where we are assuming that the initial fluid four-velocity perturbation is zero (see~\cite{GK2011} for details).

Since our initial data functions are scalable with respect to $H_a$, the results can be easily scaled to other values of the gravitational and electromagnetic emission. Numerical integration of the perturbation equations is carried out on a mesh with grid-spacing $\Delta r = R/200$ and time-step of $\Delta t/\Delta r = 0.01$. We use the same strength of the stellar magnetic field as in the black hole case, namely $B_p=10^{15}$~G, where $B_p$ is the field strength determined at $r=R$ and $\theta=0$, i.e., at the pole of star.

\begin{figure*}[htbp]
\begin{center}
\begin{tabular}{cc}
\includegraphics[width=0.45\textwidth]{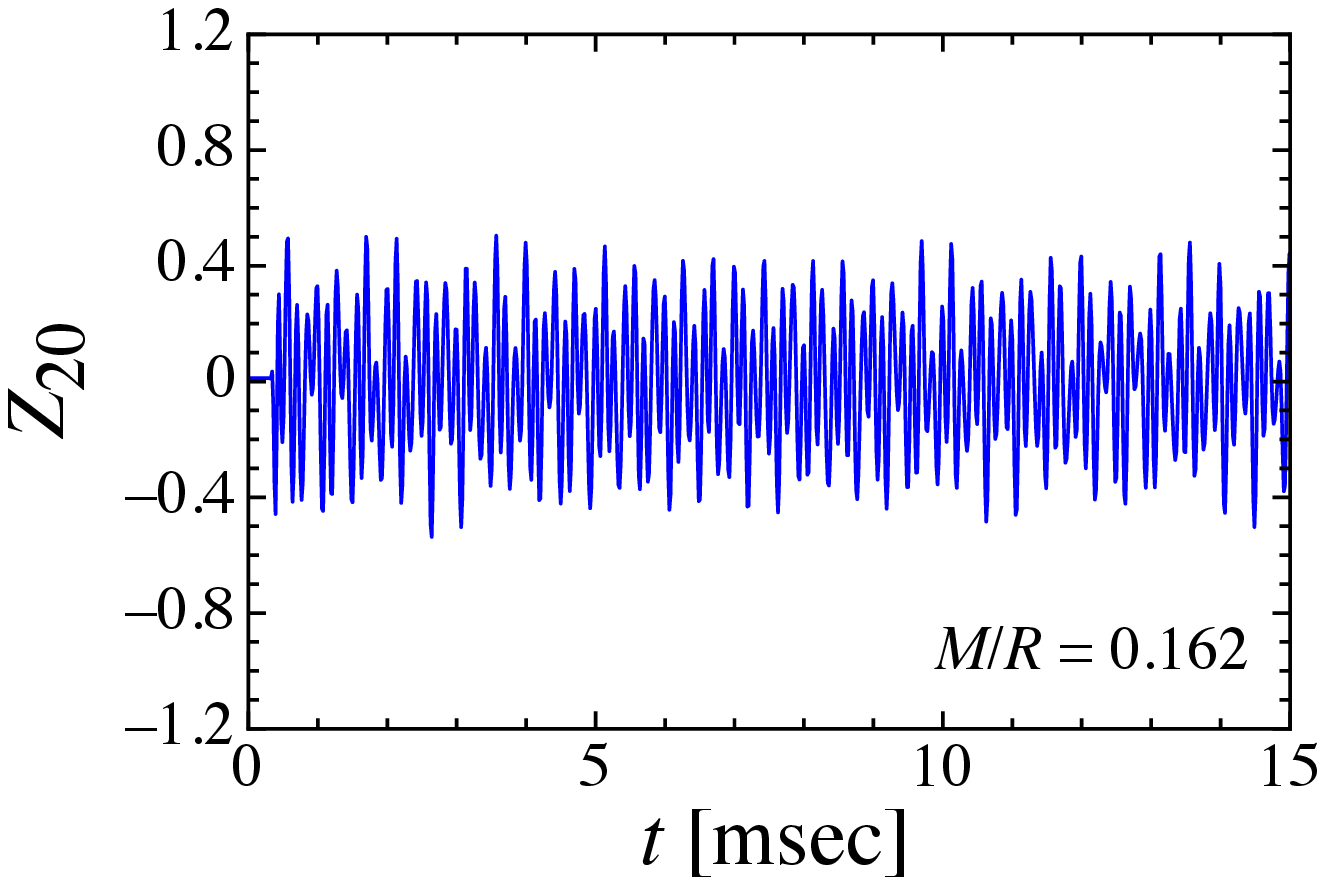} &
\includegraphics[width=0.45\textwidth]{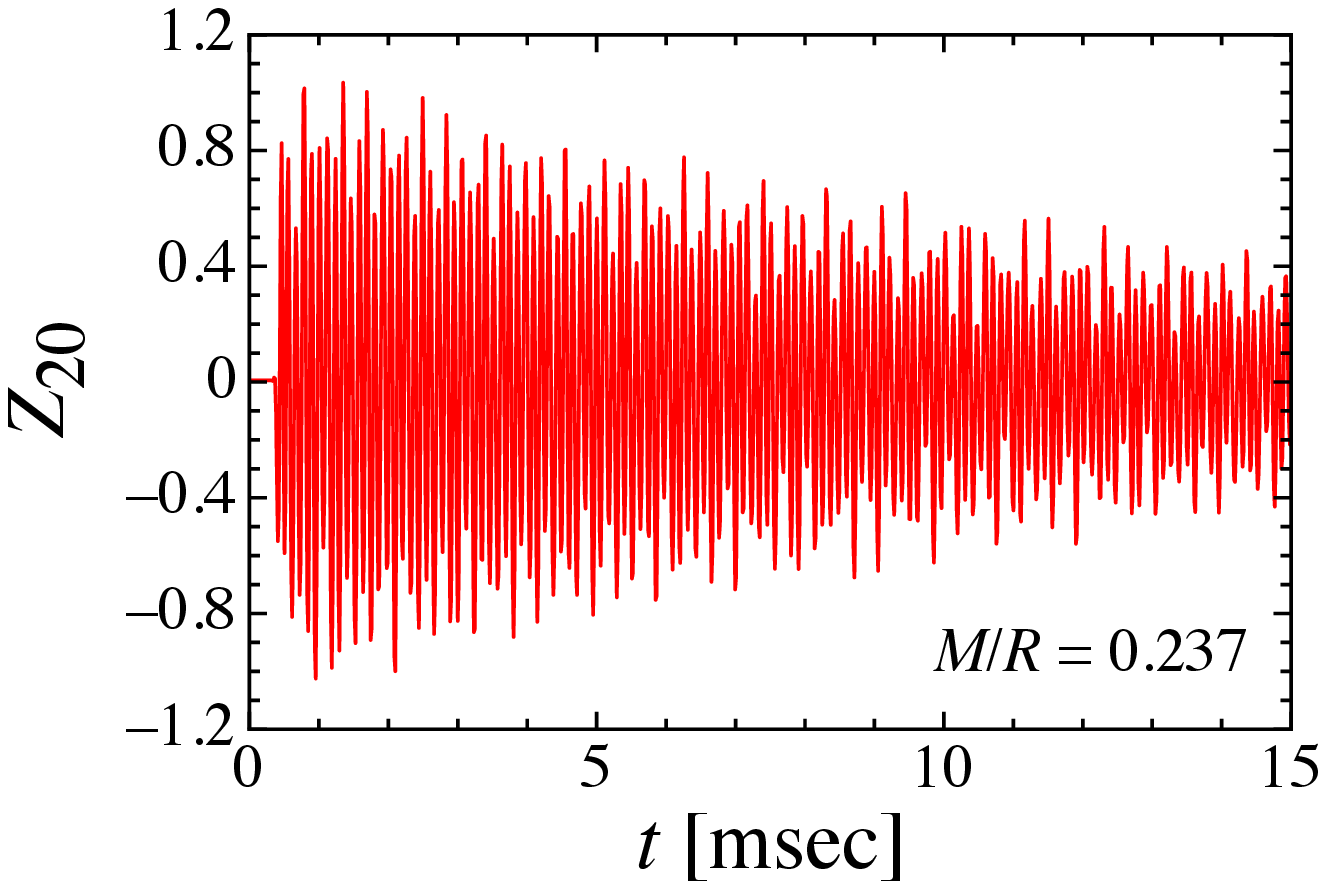}
\end{tabular}
\end{center}
\caption{
Time evolution of the Zerilli function $Z_{20}$ measured by an observer located at $r=100$ km. The left and right panels correspond to the stellar models with compactness $M/R=0.162$ and $0.237\,$, respectively.
}
\label{fig:ZZ}
\end{figure*}

\begin{figure*}[htbp]
\begin{center}
\begin{tabular}{cc}
\includegraphics[width=0.45\textwidth]{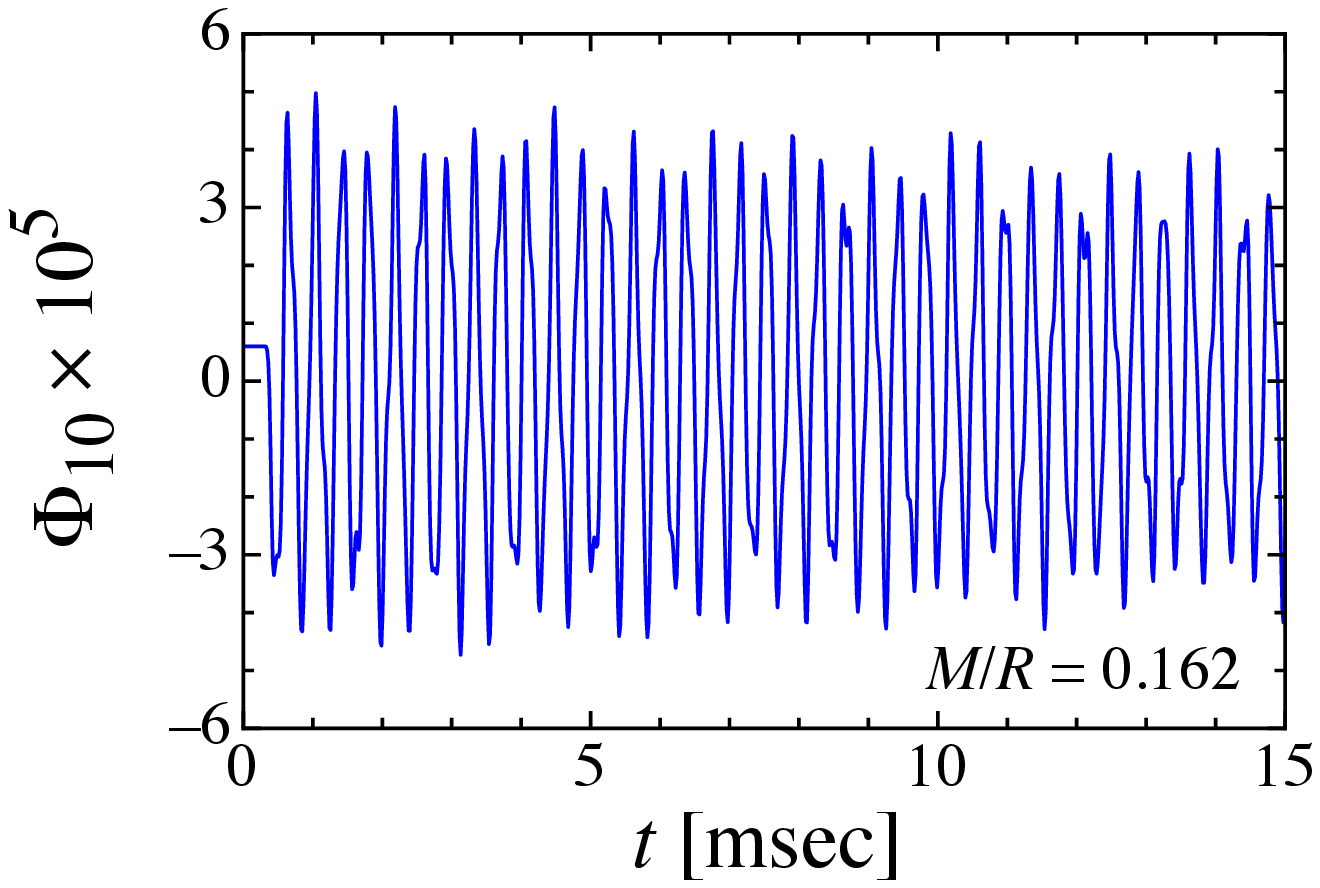} &
\includegraphics[width=0.45\textwidth]{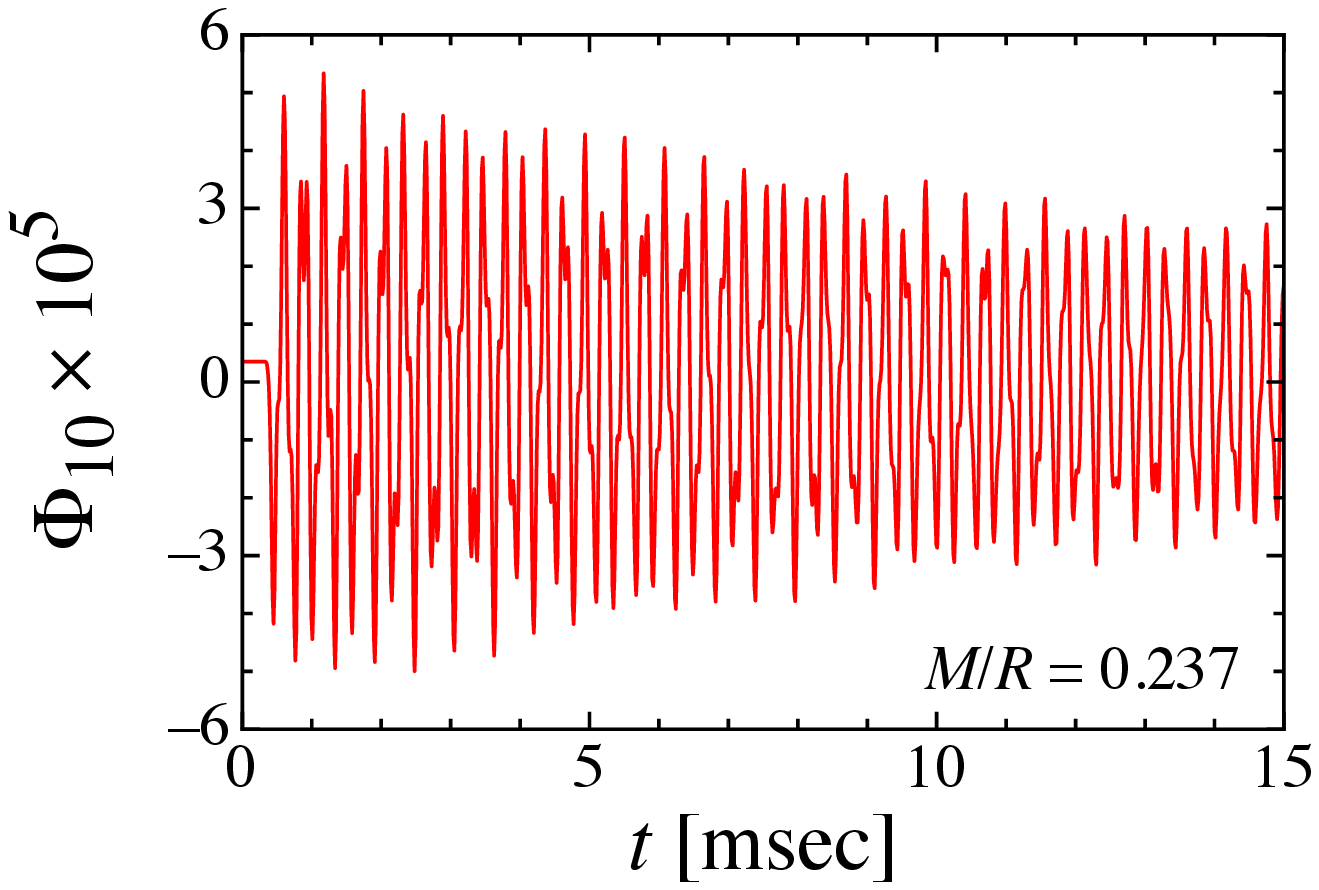}
\end{tabular}
\end{center}
\caption{
Time evolutions of the electromagnetic waves $\Phi_{10}$ driven by the polar gravitational waves, assuming the strength of the magnetic field is $B_p=10^{15}$~G. The left and right panels correspond to stellar models with compactness $M/R=0.162$ and $0.237\,$, respectively.
}
\label{fig:Phi2a}
\end{figure*}

Results of our numerical evolutions for the metric perturbations are presented in Figure~\ref{fig:ZZ} in terms of the Zerilli function $Z_{20}$, which is 
a gauge-invariant and unconstrained variable constructed from the functions $S_{20}$ and $F_{20}$.  The figure shows, in the left panel, the case of a neutron star with $(\Gamma, K) = (2,200)$, mass $M = 1.528\,M_\odot$, and compactness $M/R=0.162$, and in the right panel, the case of a neutron star with $(\Gamma, K) = (2,200)$, mass $M = 1.876\,M_\odot$, and compactness $M/R=0.237$.  A comparison of Figures~\ref{fig:GW} and~\ref{fig:ZZ} shows that the emission of gravitational radiation lasts longer for neutron stars than for black holes.  In the case of a black hole with mass $1.5\,M_\odot$, a gravitational-wave pulse lasts approximately $\sim 3\,$ msec.  It is also important to mention the role that the compactness parameter plays in the emission of gravitational waves. It is evident from Figure~\ref{fig:ZZ} that the strength of the emission increases with the compactness of the neutron star, as expected.  Finally, in Figure~\ref{fig:Phi2a} we show the evolution in time of the electromagnetic-wave emission. Not surprisingly, since the electromagnetic emission is driven by the gravitational perturbations, the profile of the waves in Figure~\ref{fig:Phi2a} mimics that of the gravitational waves showed in Figure~\ref{fig:ZZ}.

%
%

\begin{figure*}[htbp]
\begin{center}
\begin{tabular}{cc}
\includegraphics[width=0.45\textwidth]{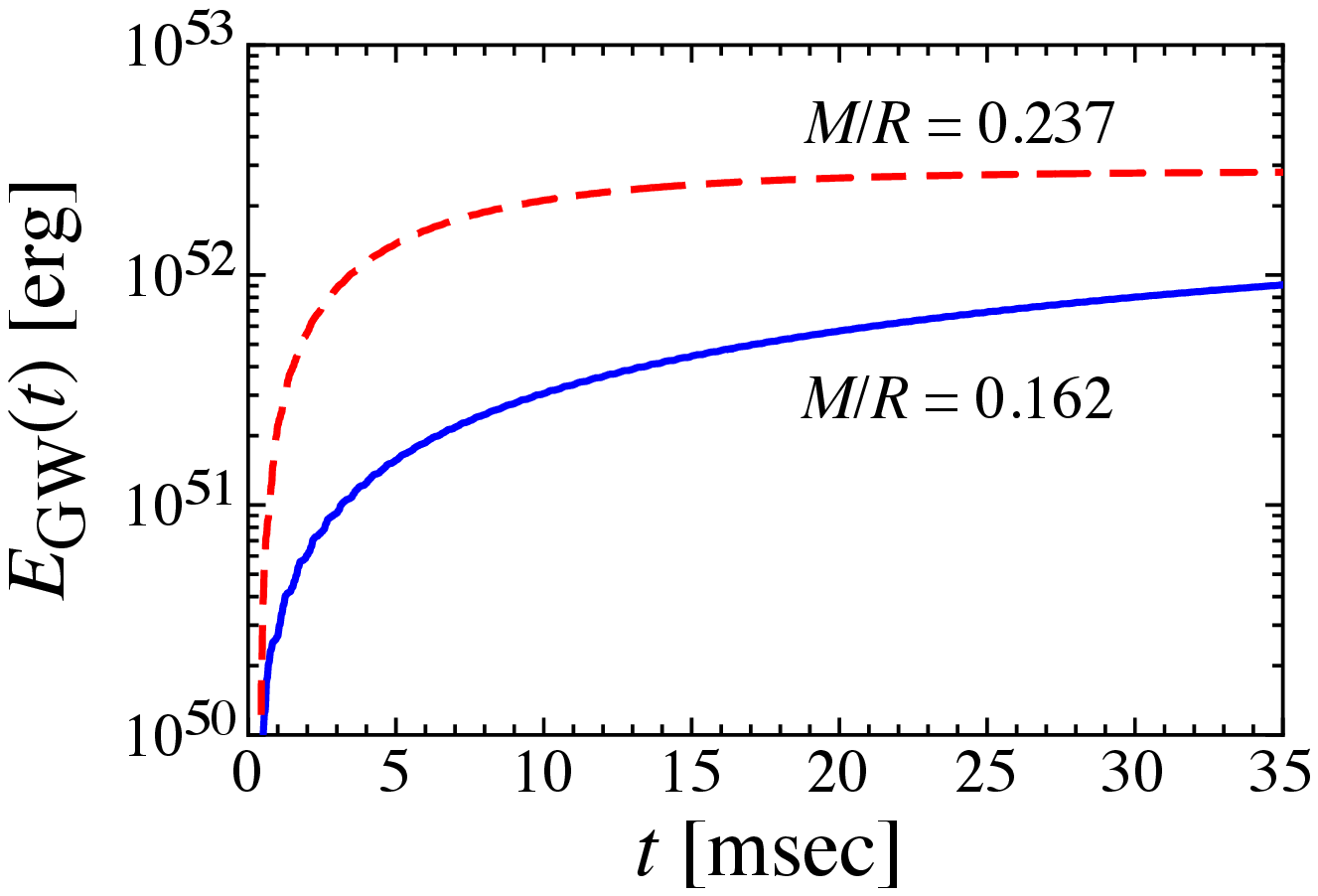} &
\includegraphics[width=0.45\textwidth]{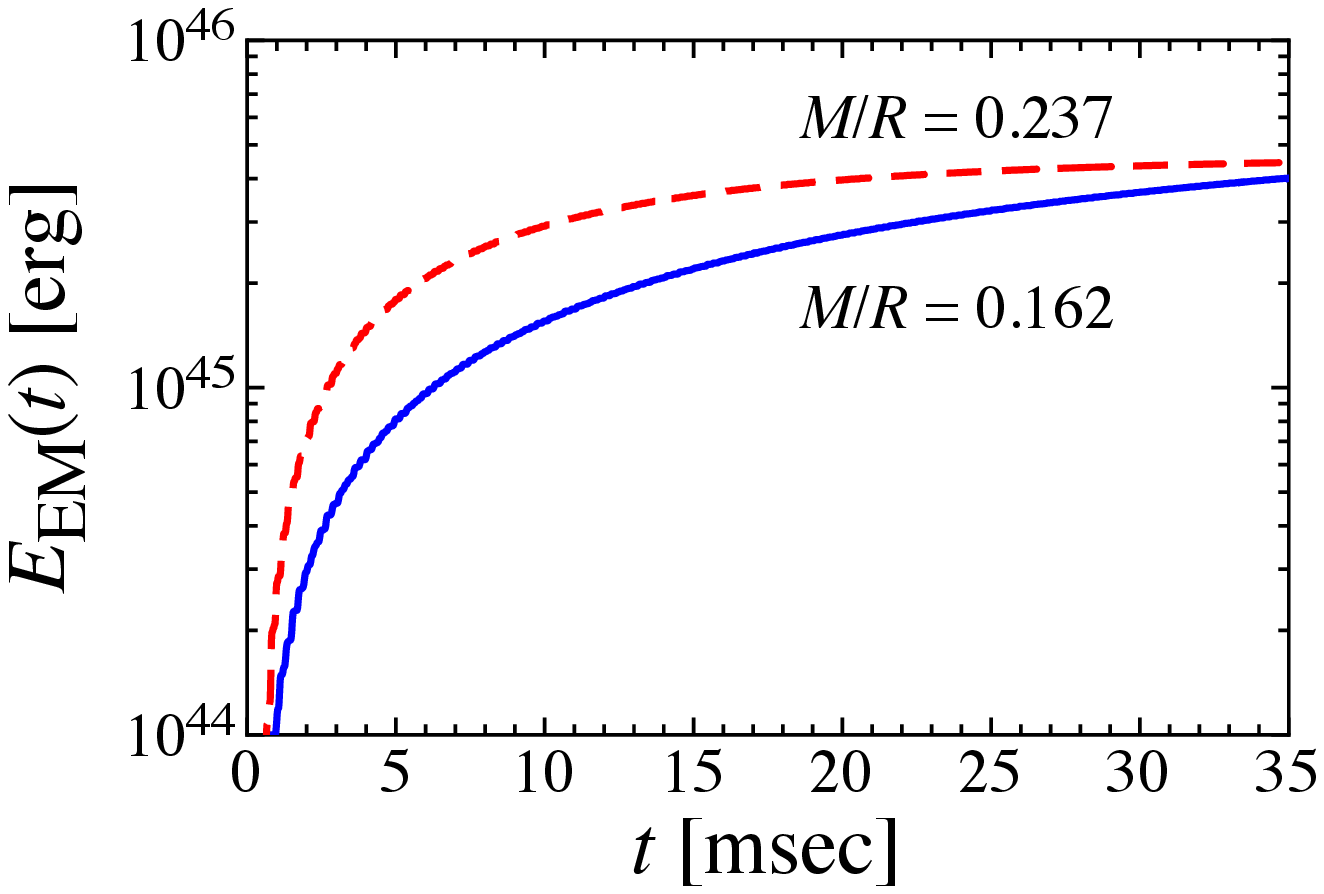}
\end{tabular}
\end{center}
\caption{
Energy radiated in gravitational (left panel) and electromagnetic (right panel) waves, assuming the strength of magnetic field is $B_p=10^{15}$~G, and for two different stellar models, one with compactness $M/R=0.237$ (dashed line) and the other one with compactness $M/R=0.162$ (solid line).
}
\label{fig:Et}
\end{figure*}

The energy emitted in gravitational and electromagnetic waves as a function of time is obtained from the luminosities: 
\begin{equation}
  E_{\rm GW}(t) = \int_0^t L_{{\rm GW}}dt \ \ {\rm and}\ \ 
  E_{\rm EM}(t) = \int_0^t L_{{\rm EM}}dt\,,
\end{equation}
where $L_{{\rm GW}}$ and $L_{{\rm EM}}$ are given in Eq.~(\ref{eq:LEM}).  The results of these computations are shown in
Figure~\ref{fig:Et}, where the dashed line corresponds to the energy emitted for the model with compactness $M/R=0.237$ and the solid line for the model with $M/R=0.162$. As expected, the stellar model with the larger compactness exhibits stronger emissions of both gravitational and electromagnetic energy. 

\begin{figure}[htbp]
\begin{center}
\includegraphics[width=0.45\textwidth]{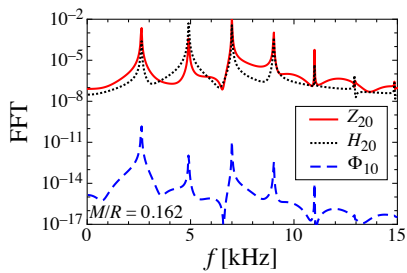}
\end{center}
\caption{
FFTs of the gravitational $(Z_{20})$, fluid $(H_{20})$, and electromagnetic $(\Phi_{10})$ perturbations for the stellar model with $M/R=0.162$.
}
\label{fig:FFT_090a}
\end{figure}

In Figure~\ref{fig:FFT_090a}, we show the FFTs of the gravitational $(Z_{20})$, fluid $(H_{20})$, and electromagnetic $(\Phi_{10})$ perturbations for the stellar model with $M/R=0.162$, using the waveforms shown in Figures~\ref{fig:ZZ} and~\ref{fig:Phi2a}, respectively. Notice that the gravitational and electromagnetic spectra have the same shape, and thus the same set of frequencies. This a natural consequence of the fact that $E_{\rm GW} \propto E_{\rm EM}$, as implied by Eq.~(\ref{eq:E}). Furthermore,  for frequencies below $12\,$kHz, the fluid perturbations have also the same set of modes although with slightly different spectrum shape. Finally, we make a comment on the spacetime modes, the so-called $w$-modes~\cite{Kokkotas1992}. They are difficult to be seen in the numerical data because such oscillations are damped out fast and in the specific problem we are dealing with the initial data favoured the excitation of the fluid modes.

\begin{figure}[htbp]
\begin{center}
\includegraphics[width=0.45\textwidth]{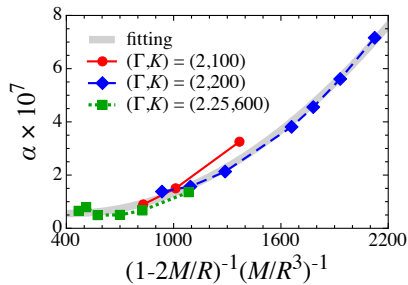}
\end{center}
\caption{
Plots of the proportionality constant $\alpha$ in Eq.~(\ref{eq:E}) as a function of ${\cal \chi}$ [see Eq.~(\ref{eq:chidef})] for various polytropic stellar models. The circles, diamonds, and squares correspond to the stellar models with $(\Gamma,K)=(2,100)$, $(2,200)$, and $(2.25,600)$, respectively. In addition to the numerical results we plot the fitting formula of Eq.~(\ref{eq:fittingformulachi}) (thick solid line).
}
\label{fig:alpha}
\end{figure}

Since for all the models we consider we have fixed the quantity $B_{15} = 1$, using Eq.~(\ref{eq:E}) we can estimate the constant $\alpha$ from the relation
$\alpha = E_{\rm EM} / E_{\rm GW}$.  In Figure~\ref{fig:alpha} we show $\alpha$ as a function of
\begin{equation}
{\cal \chi} \equiv (1-2M/R)^{-1}(M/R^3)^{-1}\,.
\label{eq:chidef}
\end{equation}
As we can see, this quantity ${\cal \chi}$ is a combination of the stellar average density $M/R^3$ and compactness $M/R$.
Each set of points in Figure~\ref{fig:alpha} represents a different stellar model, circles for $(\Gamma,K) = (2,100)$, diamonds $(\Gamma,K) = (2,200)$ and squares $(\Gamma,K) = (2.25,600)$. A fit to the points in Figure~\ref{fig:alpha} yields (thick solid line in the figure):
\begin{equation}
\alpha \times 10^7 = 0.922 - 0.903\left(\frac{{\cal \chi}}{500}\right) + 0.552 \left(\frac{{\cal \chi}}{500}\right)^2 \,.
\label{eq:fittingformulachi}
\end{equation}

%
%

%
%

%
%

\section{Conclusions}
\label{sec:conclusion}

This paper presents results of the emission of gravitational and electromagnetic waves from the strongly magnetized compact objects, extending the
study presented in Paper I~\cite{SKLS2013}. In Paper I we formulated the problem and examined in detail the case of ``electric-type" (polar) electromagnetic waves driven by ``magnetic-type'' (axial) gravitational perturbations. In the present work, we focused on the case in which ``magnetic-type" (axial) electromagnetic waves are driven by ``electric-type'' (polar) gravitational waves. As we have pointed out before, for the case of neutron stars, this situation is observationally more interesting because such type of gravitational waves are associated with fluid oscillations, and thus they will tell us about properties of the structure of the star.  

In the case of black holes, we find that the emission of ``magnetic-type" (axial) electromagnetic waves driven by ``electric-type'' (polar) gravitational perturbations is very similar to that of ``electric-type" (polar) electromagnetic waves driven by ``magnetic-type'' (axial) gravitational perturbations studied in Paper I. More specifically, we obtained the same relationship as in Paper I between the radiated energy in electromagnetic waves and the energy emitted in gravitational waves, namely $E_{\rm EM} = \alpha B_{15}^2E_{\rm GW}$. 
On the other hand, for neutron stars, we observe that the polar gravitational waves last longer than in the axial case, and thus also the radiation of  ``magnetic-type" (axial) electromagnetic waves.  
This is due to the coupling with the fluid which acts as a storage reservoir of energy which slowly leaks energy to the electromagnetic sector during a longer time period.   This process is revealed via the study of the spectra of the perturbations, where we can see that the oscillation frequencies of gravitational and electromagnetic waves are the same as those of the fluid perturbations. 
Our results also show that the larger the compactness of the star, the stronger is the coupling between the two types of perturbations leading to an enhancement in the energy transfer and the emission of electromagnetic waves.

At the same time, the emitted radiation get damped out faster for large compactness models than for those with smaller values. Unfortunately, at frequencies of a few kHz, the emitted radiation in electromagnetic waves is easily absorbed by the interstellar medium and/or the plasma around the compact objects. To be detectable, one would have to explore reprocessing mechanisms.  We are currently investigating situations/mechanism where this electromagnetic emission may be detectable and the type of information that we could extract from them.

\section*{Appendix}
\label{sec:appendix_a}

In this appendix, we present the perturbation equations for gravitational and electromagnetic waves adopted in this paper. More details can be found in Paper I \cite{SKLS2013}.
Using the well-known Regge-Wheeler gauge, the metric perturbations, $h_{\mu\nu}$, with polar parity can be decomposed into tensor spherical harmonics as 
\begin{equation}
 h_{\mu\nu} =
 \sum_{l=2}^{\infty} \sum_{m=-l}^{l}\left(
 \begin{array}{cccc}
 e^{\nu} H_{0,lm}  &  H_{1,lm} & 0 & 0 \\
 \ast & e^{\lambda} H_{2,lm}  & 0 & 0 \\
 0 & 0 & r^2 K_{lm} & 0 \\
 0 & 0 & 0 & r^2 \sin^2\theta K_{lm} \\
 \end{array}
 \right) Y_{lm}\,,
 \end{equation}
where $H_{0,lm}$, $H_{1,lm}$, $H_{2,lm}$, and $K_{lm}$ are functions of $t$ and $r$ only, and $Y_{lm}=Y_{lm}(\theta,\phi)$ denotes the scalar spherical 
harmonics on the $2-$sphere. 
On the other hand, the tensor harmonic expansion for the ``magnetic" multipoles (or axial parity) of the electromagnetic perturbations, $f_{\mu\nu}$, 
is given by
\begin{equation}
 f_{\mu\nu}^{\rm (M)} =
 \sum_{l=2}^{\infty} \sum_{m=-l}^{l}\left(
 \begin{array}{ccccccc}
 0 & & 0  & & f^{\rm (M)}_{02,lm} {\sin^{-1}\theta} \partial_{\phi}
 & &-f^{\rm (M)}_{02,lm} \sin\theta \, \partial_{\theta} \\
 0& & 0 & &f^{\rm (M)}_{12,lm} {\sin^{-1}\theta} \partial_{\phi}
 & &-f^{\rm (M)}_{12,lm} \sin\theta \, \partial_{\theta} \\
 \ast& &  \ast& & 0 && f^{\rm (M)}_{23,lm}\sin \theta \\
 \ast& & \ast && \ast && 0 \\
 \end{array}
 \right) Y_{lm}\,,
 \end{equation}
where $f_{02}^{\rm (M)}$, $f_{12}^{\rm (M)}$, and $f_{23}^{\rm (M)}$ are also functions of $t$ and $r$ only.

\subsection{Neutron-Star Background}
\label{sec:NS}

For the perturbations around the neutron star background, we adopt the formalism derived by Allen {\it et al}.~\cite{Allen1998}, 
where the perturbations are described by three coupled wave-type equations, i.e., two equations describe the perturbations of 
the spacetime and the other one describes the fluid perturbations. 
In addition to the three wave equations, there is also a constraint equation. The two wave equations for the spacetime variables are
\begin{eqnarray}
 -\frac{\partial ^2 S_{lm}}{\partial t^2} + \frac{\partial ^2 S_{lm}}{\partial r_*^2}
     &+& \frac{2 e^\nu}{r^3}\left[ 2 \pi r^3 (\rho+3p) + m - (n+1)r \right]S_{lm} \nonumber \\
     &=& -\frac{4 e^{2\nu}}{r^5} \left[\frac{(m+4\pi p r^3)^2}{r-2m} + 4\pi \rho r^3 - 3m \right] F_{lm}\,,
     \label{eq:wave_ns_m-p-S}
\end{eqnarray}
\begin{eqnarray}
 -\frac{\partial ^2 F_{lm}}{\partial t^2}
     &+& \frac{\partial ^2 F_{lm}}{\partial r_*^2}
        + \frac{2 e^\nu}{r^3} \left[2 \pi r^3 (3 \rho+ p) + m - (n+1)r \right] F_{lm} \nonumber \\
     &=& - 2\left[ 4 \pi r^2 (p+\rho) - e^{-\lambda} \right] S_{lm}
     + 8 \pi (\rho+p) {r e^{\nu}} \left(1- \frac{1}{C_s^2} \right) H_{lm} \,, \label{eq:wave_ns_m-p-F}
\end{eqnarray}
where $r_*$ is the tortoise coordinate defined as $\partial_{r_*} =e^{(\nu-\lambda)/2}\partial_r$, $n \equiv (l-1)(l+2)/2$, 
$C_s$ is the fluid sound speed, and $F_{lm}$, $S_{lm}$, and $H_{lm}$ are given by
\begin{eqnarray}
 F_{lm}(t,r) &=& rK_{lm}\,, \\
 S_{lm}(t,r) &=& \frac{e^{\nu}}{r}\left(H_{0,lm} -K_{lm}\right)\,, \\
 H_{lm}(t,r) &=& \frac{\delta p_{lm}}{\rho+p}\,.
\end{eqnarray}
On the other hand, the wave equation for the perturbations of the relativistic enthalpy, $H_{lm}$, describing the fluid perturbations, is 
\begin{eqnarray}
 &-& \frac{1}{C_s^2}\frac{\partial ^2 H_{lm}}{\partial t^2} + \frac{\partial ^2 H_{lm}}{\partial r_*^2}
        + \frac{e^{(\nu+\lambda)/2}}{r^2}\left[(m + 4\pi p r^3)\left(1-\frac{1}{C_s^2}\right) + 2 (r-2m) \right]
         \frac{\partial H_{lm}}{\partial r_*} \nonumber \\
     &+& \frac{2 e^\nu}{r^2} \left[ 2 \pi r^2 (\rho+p)\left(3 + \frac{1}{C_s^2} \right) - (n+1) \right] H_{lm}
         \nonumber \\
     &=& (m+4 \pi p r^3)\left(1-\frac{1}{C_s^2}\right) \frac{e^{(\lambda-\nu)/2}}{2 r}
         \left(\frac{e^\nu}{r^2}\frac{\partial F_{lm}}{\partial r_*} - \frac{\partial S_{lm}}{\partial r_*}
         \right) \nonumber \\
     &+& \left[\frac{(m+4\pi p r^3)^2}{r^2(r-2m)}\left(1+\frac{1}{C_s^2}\right) - \frac{ m+4\pi p r^3}{2 r^2} 
         \left(1-\frac{1}{C_s^2}\right)-4\pi r (3p+\rho) \right] S_{lm} \nonumber \\
     &+& \frac{e^\nu}{r^2}\left[\frac{2(m+4\pi p r^3)^2}{r^2(r-2m)}\frac{1}{C_s^2}
         - \frac{ m+4\pi p r^3}{2 r^2}\left(1-\frac{1}{C_s^2} \right) -4\pi r (3p+\rho)\right] F_{lm}\,.
 \label{eq:wave_ns_m-p-H}
\end{eqnarray}
This third wave equation is valid only inside the star.  On the other hand, the first two wave equations are simplified considerably
outside the star, and they can be reduced to a single wave equation, i.e., the Zerilli equation (see~\cite{Allen1998} and \S\ref{sec:BH}).
Finally, the Hamiltonian constraint,
\begin{eqnarray}
 \frac{\partial ^2 F_{lm}}{\partial r_*^2}
     &-& \frac{e^{(\nu+\lambda)/2}}{r^2} \left( m + 4 \pi r^3 p \right) \frac{\partial F_{lm}}{\partial r_*}
         + \frac{e^\nu}{r^3} \left[ 12\pi r^3 \rho - m - 2(n+1)r \right] F_{lm}
         \nonumber \\
     &-& r e^{-(\nu+\lambda)/2}\frac{\partial S_{lm}}{\partial r_*} + \left[8\pi r^2(\rho+p) -(n+3)
         + \frac{4m}{r} \right] S_{lm} \nonumber \\
     &+& \frac{8 \pi r}{C_s^2} e^\nu (\rho+p) H_{lm} = 0\,,
 \label{Hamilton}
\end{eqnarray}
can be used for setting up initial data and monitoring the evolution of the coupled system.

Finally, the perturbation equation for the  ``magnetic-type" (axial) electromagnetic perturbations outside the neutron star is
\begin{equation}
 \frac{\partial^2 \Phi_{10}}{\partial t^2} - \frac{\partial^2 \Phi_{10}}{\partial r_*^2}
     + \frac{2}{r^2}e^{\nu}\Phi_{10} = {\cal S}_{20}^{\rm (M)}\,, \label{eq:wave-magnetic}
\end{equation}
where $\Phi_{lm} \equiv f_{23,lm}^{\rm (M)}$ and
\begin{eqnarray}
 {\cal S}_{20}^{\rm (M)} &=& -\frac{2}{\sqrt{15}}e^{\nu}\bigg[\left(2B_2 + r {B_2}'\right)r^2S_{20} \nonumber \\
     &+& \left(e^{\nu}B_2 + re^{\nu}{B_2}' - \frac{2}{r}B_1\right)F_{20}
     + rB_2 \frac{\partial F_{20}}{\partial r_*}\bigg]\,. \label{eq:source-magnetic}
\end{eqnarray}

\subsection{Black-Hole Background}
\label{sec:BH}

The ``magnetic-type" (axial) electromagnetic perturbations driven by the gravitational perturbations on a black hole background 
are also governed by the same equation as the one derived for the exterior region of neutron stars [see Eq.~(\ref{eq:wave-magnetic})], 
that is
\begin{eqnarray}
 \frac{\partial^2 \Phi_{10}}{\partial t^2} - \frac{\partial^2 \Phi_{10}}{\partial r_*^2}
     + \frac{2}{r^2}e^{\nu}\Phi_{10} = {\cal S}_{20}^{\rm (M)}\,.
\end{eqnarray}
The source term of this wave equation, ${\cal S}_{20}^{(\rm M)}$, has the same form as in Eq.~(\ref{eq:source-magnetic}). 
On the other hand, and is well-known, the spacetime metric perturbations are described by a single equation, the Zerilli equation
\begin{equation}
 \frac{\partial^2 Z_{lm}}{\partial t^2}-\frac{\partial^2 Z_{lm}}{\partial r_*^2} + V_Z(r)Z_{lm} = 0
\end{equation}
\begin{equation}
 V_{Z}(r) = \frac{2e^{\nu}\left[n^2(n +1)r^3 + 3Mn^2 r^2 + 9M^2n r + 9M^3\right]}
     {r^3(nr + 3M)^2}\,.
\end{equation}
In the same way as in the case of the neutron star background, one can also adopt $F_{lm}$ and $S_{lm}$ as the perturbation variables 
for the spacetime oscillations. Actually, one can obtain the variables $F_{lm}$ and $S_{lm}$ 
from the Zerilli function $Z_{lm}$ using the following expressions:
\begin{eqnarray}
 F_{lm} &=& r\frac{dZ_{lm}}{dr_*}+\frac{n(n+1)r^2+3 n Mr+6M^2}{r(n r+3M)}Z_{lm}\,, \\    \label{eq:ZF}
 S_{lm} &=& \frac{1}{r}\frac{dF_{lm}}{dr_*}-\frac{(n+2)r-M}{r^3}F_{lm}+\frac{(n+1)(n r+3M)}{r^3}Z_{lm}\,.  \label{eq:ZS}
\end{eqnarray}

\begin{acknowledgements}
H.S. is grateful to Y. Sekiguchi for valuable comments.
H.S., K.K., and P.L. acknowledge the support and hospitality of the YITP (Kyoto) during the 
workshop ``YKIS2013", where a significant part of the work has been done.
This work was supported by the German Science Foundation (DFG) via SFB/TR7, 
by Grants-in-Aid for Scientific Research on Innovative Areas through No.\ 24105001, and No.\ 24105008 provided by MEXT, 
by Grant-in-Aid for Young Scientists (B) through No.\ 24740177 provided by JSPS,
by the Yukawa International Program for Quark-hadron Sciences, and 
by the Grant-in-Aid for the global COE program ``The Next Generation of Physics, Spun from Universality and Emergence" from MEXT.
P.L. is supported by NSF grants 1205864, 1205864, 0903973, 0855423.
C.F.S. is supported by contracts AYA-2010-15709 and FIS2011-30145-C03-03 of 
the Spanish Ministry of Science and Innovation, and contract 2009-SGR-935 of AGAUR (Generalitat de Catalunya).

\end{acknowledgements}




\end{document}